\documentclass[aps,prd,twocolumn,groupedaddress,superscriptaddress,letterpaper]{revtex4-1}
\usepackage{booktabs}
\usepackage{graphicx}  
\usepackage{color}     
\usepackage{amsmath}   
\usepackage{amsfonts}   
\usepackage{setspace}  
\usepackage{dcolumn}   
\usepackage{enumitem}  
\usepackage{bm}        
\usepackage[usenames,dvipsnames]{xcolor}
\usepackage{geometry}

\begin{document}

\title{The NDL Equation of State for Supernova Simulations}
\author{Matthew Meixner}
\email[]{mmeixner@nd.edu}
\affiliation{Center for Astrophysics, Department of Physics, University of Notre Dame, Notre Dame, IN 46556}

\author{J. Pocahontas Olson}
\email[]{jspeare@nd.edu}
\affiliation{Center for Astrophysics, Department of Physics, University of Notre Dame, Notre Dame, IN 46556}

\author{Grant Mathews}
\email[]{gmathews@nd.edu}
\affiliation{Center for Astrophysics, Department of Physics, University of Notre Dame, Notre Dame, IN 46556}

\author{N. Q. Lan}
\email[]{nquynhlan@hnue.edu.vn}
\affiliation{Hanoi National University of Education, 136 Xuan Thuy, Hanoi, Vietnam}
    
\author{H. E. Dalhed}
\affiliation{Lawrence Livermore National Laboratory,
    Livermore, CA, 94550}

\date{\today} 

\begin{abstract}
We present an updated and improved equation of state (which we call the NDL~EoS) for use in neutron-star structure and supernova simulations. This EoS is based upon a framework originally developed by Bowers \& Wilson, but there are numerous changes.
Among them are:
(1) a reformulation in the context of density functional theory;
(2) the possibility of the formation of material with a net proton excess ($Y_e > 0.5$);
(3) an improved treatment of the nuclear statistical equilibrium and the transition to heavy nuclei as the density approaches nuclear matter density;
(4) an improved treatment of the effects of pions in the regime above nuclear matter density including the incorporation of all the known mesonic and baryonic states at high temperature;
(5) the effects of 3-body nuclear forces at high densities; and
(6) the possibility of a first-order or crossover transition to a QCD chiral symmetry restoration and deconfinement phase at densities above nuclear matter density.  
This paper details the physics of, and constraints on, this new EoS and describes its implementation in numerical simulations.
We show comparisons of this EoS with other equations of state commonly used in supernova collapse simulations. 
\end{abstract}

\maketitle
\section{Introduction}
To describe the hydrodynamics of compact matter; be it in heavy-ion nuclear collisions, supernovae or neutron stars; an equation of state (EoS) is needed to relate the physics of the state variables.  In supernovae the EoS determines the dynamics of the collapse and the outgoing shock, and determines whether the remnant ends up as a neutron star or a black hole.  In a neutron star, it determines the maximum mass, mass-radius relationship, internal composition, cool-down time and dynamics of neutron star mergers. 

At present, only a few hadronic EoSs are commonly employed that cover large enough ranges in density, temperature and electron fraction to be of use in core-collapse supernova simulations.  These EoSs are usually in the form of a multi-dimensional table with three independent variables (e.g. density, temperature and electron~fraction).  

The two most commonly used equations of state in astrophysical simulations are the EoS of Lattimer \& Swesty (LS91)~\cite{LS91} and that of Shen~et.~al.~(Shen98)~\cite{Shen98a, Shen98b}.  The former utilizes a non-relativistic parameterization of nuclear interactions in which nuclei are treated as a compressible liquid drop including surface effects. The latter is based upon a Relativistic Mean Field~(RMF) theory using the TM1 parameter set in which nuclei are calculated in a Thomas-Fermi approximation. Recently, Shen~et.~al.~\cite{Shen2011} released a updates of the Shen98 EoS table. The first update, EoS2, increased the number of temperature points as well as switching to a linear grid spacing in $Y_p$. In the second update, EoS3, the effects of $\Lambda$ hyperons were taken into account. It should be noted that several extensions to the Shen98 table have also been proposed, either by the implementation of hyperons~\cite{Ishizuka2008} or, of particular relevance to the present work, including a mixed phase transition to a quark gluon plasma~\cite{Fischer2011,Sagert}. 

Over the last several years much progress has been made on other formulations of the supernova EoS. Several new EoS tables, each based upon RMF models, have been introduced~\cite{GShenNL3, Steiner2012}.  The new hadronic tables of Shen~et.~al.~\cite{GShenNL3} are based upon a virial expansion and two different RMF interactions in the Hartree approximation. The first table includes the NL3~\cite{Lalazissis} parameter set while the second table is parameterized by the FSUgold parameters~\cite{GShenFSUgold}.  The new EoS of Hempel~et.~al.~\cite{Hempel2010} is described by an RMF in nuclear statistical equilibrium (NSE) for an ensemble of nuclei and interacting nucleons.  Steiner~et.~al.~\cite{Steiner2012} also constructed several new EoSs to match recent neutron star observations. In these nucleonic matter was parameterized with a new RMF model that treated nuclei and non-uniform matter with the statistical model of Hempel~et.~al.~\cite{Hempel2010}.

In this work we describe a new Notre Dame-Livermore (NDL) EoS that we make publicly available. This EoS evolves from the original Livermore formulation~\cite{Bowers82,WilsonMathews}. This NDL~EoS, is consistent with known experimental nuclear matter constraints and recent mass and radius measurements of neutron stars. 

Below nuclear matter density, the conditions for NSE are imposed in the NDL EoS above a temperature of $T \approx 0.5$ MeV.  Below this temperature the nuclear matter is a approximated by a nine element reaction network which must be evolved dynamically. Above this temperature, the nuclear constituents are represented by free nucleons, alphas and a single ``representative'' heavy nucleus. The high density phase of the EoS is treated with a parameterized Skyrme energy density functional that utilizes a modified zero range 3-body interaction.  The effects of pions and of the mesonic and baryonic resonances on the state variables at high densities are also included as well as the consequences of a phase transition to a QGP. For the NSE and high density regime, the EoS is provided in tabular form covering the necessary ranges in density, temperature and $Y_e$ needed for use in astrophysical simulations. Below nuclear saturation where NSE cannot be applied, numerical routines are available to the user \footnote{The NDL~EoS is available upon request from the authors.}.


\section{The NDL Equation of State}
Depending upon the density and temperature there are a variety of matter components that contribute significantly to the equation of state  during various epochs of supernova collapse and the interiors of neutron stars. These include photons, electrons, positrons, neutrinos, mesons, all the 215 known mesonic and baryonic states~\cite{PDGwhite}, free neutrons, protons, and atomic nuclei, and even the possibility of a 
crossover to a quark gluon plasma.  At low density and high temperatures we assume a meson gas; consisting of thermally created, pair-produced mesons with zero chemical potential.  In the high density limit, but low temperatures, pions are constrained by chemical equilibrium among the neutrons, protons and the other baryonic states.  Baryons are assumed to have the same, non-zero chemical potential, such that baryon number is conserved.  The inclusion of the additional mesonic and baryonic states is yet another improvement appearing in this updated EoS.

Since the material is optically thick to photons, one can include photons along with matter particles in the equation of state. The NDL~EoS is divided into four regimes: 
\begin{enumerate}
	\item Baryons below nuclear matter density and not in NSE; 
	\item	Baryons below nuclear matter density and in NSE; 
	\item Hadronic matter above saturation density including pions; and 
	\item A first order or cross-over phase transition to quark gluon plasma.
\end{enumerate}

The electrons and positrons are approximated as a uniform background and are treated as a non-interacting ideal Fermi-Dirac gas.  Photons are approximated as a black-body and thus are given by the usual Stefan-Boltzmann law. Neutrinos, however, are not necessarily confined and must be transported dynamically. In supernova simulations most matter except neutrinos can be assumed to be in local thermodynamic equilibrium (one temperature in a zone) but not necessarily in chemical equilibrium (i.e. the weak reactions have not necessarily equilibrated).  The independent variables generally chosen for the equation of state are then the temperature $T$, the matter rest-mass density $\rho$, and the net charge per baryon $Y_e = n_e / n_B$.  The previous formulation required that $Y_e < 0.5$, but we have removed that restriction in this new version.

\subsection{Baryons Below Saturation \\and not in NSE}
Below nuclear saturation density and above a temperature of T $\approx$ 0.5 MeV we can assume that NSE is valid. Below this temperature the isotopic abundances must be evolved dynamically. To achieve this, the nuclear constituents are approximated by a 9 element nuclear burn network consisting of n, p, 
$^4$He, $^{12}$C, $^{16}$O, $^{20}$Ne, $^{24}$Mg, $^{28}$Si, $^{56}$Ni~\cite{Bowers}.  The free energy per baryon is taken to be the sum of contributions from an ideal gas $F_g$~[Eq.~(\ref{eqn:Freegas})] and a coulomb correction $F_c$~[Eq.~(\ref{eqn:Coulomb})]. 
The ideal gas contribution is simply,
\begin{equation}
F_g = \sum_i\left[\frac{T}{A_i}\text{ln}\left(\frac{X_in\frak{\cal A}}{T^{3/2}A_i^{5/2}}\right)\right].
\label{eqn:Freegas}
\end{equation}
The relevant variables here are the nuclear mass fraction $X_i$, the local baryon number density $n$, the temperature $T$, and the atomic mass number $A_i$. The index $i$ runs over the entire reaction network, and ${\cal A}$ is the thermal wavelength per baryon given by
\begin{equation}
\frak{\cal A} = \frac{8\pi^3}{e\left(2\pi m_B\right)^{3/2}}.
\label{eqn:thermalwave}
\end{equation}
[Note, that natural units ($\hbar = c = k = 1$) have been adopted here and throughout this manuscript.]

The Coulomb contribution is given by
\begin{equation}
	\label{eqn:Coulomb}
	F_c = -\frac{1}{3}n^{1/3}e^2\langle A\rangle^{2/3}Y_e^2.
\end{equation}
From these relations the baryonic pressure and energy per unit mass can be calculated from the ideal gas thermodynamic relations.
\begin{equation}
P_m = Tn\sum_i \frac{X_i}{A_i}-\frac{1}{3m_B}n^{4/3}e^2\langle A \rangle^{2/3}Y_e^2,
\end{equation}
\begin{equation}
\epsilon_{m} = \frac{3}{2}\frac{T}{m_B} \sum_i \frac{X_i}{A_i} - \frac{1}{3m_B}n^{1/3}e^2\langle A \rangle^{2/3}Y_e^2.
\end{equation}

\subsection{Baryons Below Saturation and in NSE}

When NSE is valid, the baryonic nuclear material is approximated as consisting of a 4 component fluid of free protons, neutrons, alpha particles and an average heavy nucleus. This formulation is reasonably accurate and convenient in that it leads to fast analytic solutions for the NSE. One should exercise caution, however,~\cite{Bowers82} when considering detailed thermonuclear burning or a precise value of $Y_e$ in NSE is desired. In such cases an extended NSE network should be employed.

In the absence of weak interactions the neutron and proton mass fractions are constrained by charge conservation (i.e. constant electron fraction $Y_e$),
\begin{equation}
    \sum_i Y_iX_i = Y_e 	~,
\end{equation}
and baryon conservation, i.e.
\begin{equation}
    \sum_i X_i    = 1 	~.
\end{equation}

The thermodynamic quantities are determined from the Helmholtz-free energy per baryon which is given as a sum of the various constituents,
\begin{equation}
f = f_n + f_p + f_\alpha + f_A      .
\end{equation}
The constituent free energies can be written analytically as 
\begin{widetext}
\begin{align}
    \label{Eq:fp}
    f_p &= X_B Y_p \biggl\{ \epsilon_{p 0} W + \epsilon_N(1 - W)                  
        + \frac{3}{2} kT \biggl[ \sqrt{1 + \zeta_p^2} - \ln \biggl(
            \frac{1 + \sqrt{1 + \zeta_p^2}}{\beta \zeta_p} \biggr) \biggr] \biggr\}  , \\
    \label{Eq:fn}
    f_n &= X_B Y_n \biggl\{ \epsilon_{n 0} W + \epsilon_N(1 - W)                    
           + \frac{3}{2} kT \biggl[ \sqrt{1 + \zeta_n^2} - \ln \biggl(
            \frac{1 + \sqrt{1 + \zeta_n^2}}{\beta \zeta_n} \biggr) \biggr] \biggr\}  , \\
    \label{Eq:falpha}
    f_\alpha &= X_\alpha  \biggl\{ \epsilon_{\alpha 0} W + \epsilon_N(1 - W)         
                     +  \frac{1}{4} kT \text{ln}\biggl(\frac{ X_\alpha \rho \alpha }{ T^{3/2} 4^{5/2}}\biggr) \biggr\}  ,  \\
    \label{Eq:fA}
    \begin{split}
    f_{\langle A \rangle}
        &= X_A  \biggl[ -\frac{1}{3} \biggl( \frac{\rho}{m_B}\biggr)^{1/3} e^2 \langle A \rangle^{2/3} Y_A^2 + S_E (Y_{Fe} - Y_A)^2  + \epsilon_N(1 - W) + \frac{3}{4} \rho^{4/3} Y_A^2 b(Y_e) \\
            &\qquad \quad+ \frac{kT}{A} \text{ln}\biggl(\frac{ X_A \rho \alpha}{g_A T^{3/2} A^{5/2}} \biggr) \biggr]   ,
      \end{split}
\end{align}
\end{widetext}
where the various terms in \mbox{Eqs.~(\ref{Eq:fp}) - (\ref{Eq:fA})} are defined as follows:
\begin{gather}
    \langle A \rangle = 194.0 (1 - Y_e)^2(1 + X + 2X^2 + 3 X^3)  ,    \\
  \intertext{is the density dependent mass of the average heavy nucleus, expanded in terms of the density parameter $X$, defined by}
    X \equiv \biggl( \frac{ \rho }{ 7.6 \times 10^{13} ~{\rm g~cm^{-3}}} \biggr)^{1/3} .
\end{gather}
$X_B$ is the free baryon mass fraction while $X_\alpha$ and $X_A$ are the mass fractions of $^4$He and the average heavy nucleus in obvious notation.  The quantities $Y_p$ and $Y_n$ are the relative number fractions of free baryons in protons or neutrons, respectively.  Thus, $Y_p + Y_n = 1$.  The quantity $Y_A$ is the average $Z/A$ for heavy nuclei. The quantity $W$ in \mbox{Eqs.~(\ref{Eq:fp}) - (\ref{Eq:fA})} is a weighting factor that interpolates between the low-density and high-density regimes. It is defined by $W \equiv (1 - \rho/\rho_N)^2$. The transition from subnuclear to supra-nuclear density is expected to be continuous.  The reason for this is that, as the density increases, the equilibrium continuously shifts to progressively heavier nuclei. 

When a relativistic Thomas-Fermi representation of the electrons is evaluated at subnuclear density, the electron energy is lowered by more than $\sim$1~MeV~\cite{WilsonMathews}.  The electrostatic nuclear energy also increases in magnitude. Similarly the transformation of nuclei from spheres to other more exotic shapes (e.g. pasta nuclei, etc.)~\cite{PastaRef1,PastaRef2} also lowers the energy of the medium by about 1 MeV.  The net result is that the pressure and energy are smooth functions of density near the nuclear saturation density.  Hence, the weighting factor is chosen to approximate this smooth transition.

The normal $^{56}$Fe ground state is taken as the zero of binding energy.  This is unlike most other equations of state for which the zero point is chosen relative to dispersed free nucleons. The reason for the choice made here is that it avoids the numerical complication of negative internal energies in the hydrodynamic state variables at low temperature and density due to the binding energy of nuclei. The energy per nucleon required to dissociate $^{56}$Fe into free nucleons is $\epsilon_{p 0} = 8.37$~MeV for protons, while for neutrons it is $\epsilon_{n 0} = 9.15$~MeV.

The quantity $\rho_N $ is the density at which nuclear matter becomes a uniform sea of nucleons. This was found by fitting the saturation density of nuclear matter [i.e. $P_M(\rho, T=0, Y_e) = 0$] as a function of $\rho$ and $Y_e$. The zero-temperature result was chosen to simplify the problem
of making a smooth transition between the three equation of state regimes. The result is
\begin{equation}
    \rho_N = 2.66 \times 10^{14} \left[1 - (1 - 2 Y_e)^{5/2} \right]  	~.
\end{equation}

The quantities $\zeta_n$ and $\zeta_p$  in Eqs ({\ref{Eq:fp}) and (\ref{Eq:fn}) are a measure of the degeneracy
of the free baryons.  They are defined by
\begin{equation}
    \zeta_n = \frac{{\cal B}(\rho Y_n X_B)^{2/3}}{kT} ;
    \quad
    \zeta_p = \frac{{\cal B}(\rho Y_p X_B)^{2/3}}{kT}  ,
\end{equation}
where the quantity ${\cal B}(\rho Y_i X_B)^{2/3}$ is the energy per baryon of a zero-temperature, non-relativistic ideal fermion gas and the constant ${\cal B}$ is
\begin{equation}
    {\cal B} = \frac{3}{10} \biggl(\frac{3}{8 \pi} \biggr)^{2/3} \frac{h^2}{m_B^{5/3}} ~.
\end{equation}

The dimensionless constant $\beta$ appearing in Eqs. \eqref{Eq:fp} and \eqref{Eq:fn} is determined such that the translational part of $f_p$ and $f_n$ reduces to the correct non-degenerate limit ($T\rightarrow \infty$, $\zeta_i \rightarrow 0$).
That is,
\begin{align}
\begin{split}
    \frac{3}{2} &kT \biggl[ \sqrt{1 + \zeta_n^2} - \ln{\biggl(\frac{1 + \sqrt{1 + \zeta_n^2} }{\beta \zeta_n} \biggr)} \biggr] \\
        &\rightarrow kT \ln{\biggl(\frac{X_B \rho Y_i {\cal A}}{T^{3/2}} \biggr)} .
\end{split}
\end{align}
This requirement implies
\begin{equation}
    \beta = \biggl(\frac{{\cal A}}{2} \biggr)^{2/3}\biggl(\frac{3}{e {\cal B}} \biggr) = 0.781   ~,
\end{equation}
where ${\cal A}$ is the thermal wavelength per baryon given in Eq.~(\ref{eqn:thermalwave}).

The function $b(Y_e)$ in Eq.~(\ref{Eq:fA}) is determined by the condition that the Coulomb contribution
to the pressure at $\rho = \rho_N$ be canceled by the term proportional to $b(Y_e)$.  This requires,
\begin{equation}
    b(Y_e) = \frac{ e^2}{18} \biggl(\frac{\langle A \rangle^2}{m_B} \biggr)^{\frac{1}{3}} \biggl[ \frac{1}{\rho_N} + 2 \biggl( \frac{\partial \ln{\langle A \rangle}}{\partial \rho} \biggr)_{\rho_N} \biggr]     .
\end{equation}

The expression for the statistical weight of the heavy nucleus $g_A$ appearing in Eq.~(\ref{Eq:fA}) is taken to be
\begin{gather}
    \begin{split}
        \frac{1}{A} \ln{g_A} &= \frac{3}{2} \Biggl\{ \biggl[ 1 - \sqrt{1+ \biggl( \frac{T}{T_S} \biggl)^2} \biggr] \frac{T}{T_S}  \\
                             &\quad + \ln{ \biggl[ \frac{T}{T_S} + \sqrt{1 + \biggl( \frac{T}{T_S} \biggl) }  \biggr] } \Biggr\} ,
    \end{split}  \\
 \intertext{where}
    T_S = (8 \text{ MeV}) \biggl( 1 + 2 \frac{\rho}{\rho_N} \biggr)   .
\end{gather}
In Eq.~(\ref{Eq:fA}) the constant $S_E =120$ MeV is derived for a symmetry energy of 30.4 MeV per nucleon (see below).  The constant $Y_{Fe} = 0.464$ is the fraction of protons in $^{56}$Fe.

The chemical potentials are found from the free energy as
\begin{gather}
  \label{Eq::ChemPots}
    \mu_n = \biggl( \frac{\partial F}{\partial X_B} - \frac{Y_p}{X_B} \frac{\partial F}{\partial Y_p} \biggr)  ,\\
    \mu_p = \biggl( \frac{\partial F}{\partial X_B} + \frac{Y_n}{X_B} \frac{\partial F}{\partial Y_p} \biggr)  ,\\
    \mu_\alpha =  4 \biggl(\frac{\partial F}{\partial X_\alpha}\biggr) ,\\
    \mu_{nA} = \biggl( \frac{\partial F}{\partial X_A} - \frac{Y_A}{X_A} \frac{\partial F}{\partial Y_A} \biggr)  ,\\
    \mu_{pA} = \biggl( \frac{\partial F}{\partial X_A} + \frac{(1 - Y_A)}{X_A} \frac{\partial F}{\partial Y_A} \biggr)  ,
\end{gather}
where $\mu_p$, $\mu_n$ and $\mu_\alpha$ are the chemical potentials of free protons, neutrons, and alpha particles.  The quantities $\mu_{nA}$ and $\mu_{pA}$ are the chemical potentials of neutrons and protons within heavy nuclei.  These quantities are related by the Saha equation:
\begin{gather}
    2 \mu_n + 2 \mu_p = \mu_\alpha   \\
    2 \mu_{nA} + 2 \mu_{pA} = \mu_\alpha  \\
    \mu_{nA} - \mu_{pA} =  \mu_n - \mu_p =  \hat{\mu}.
\end{gather}
In the original Livermore formulation~\cite{Bowers82,WilsonMathews}, an analytical approximation was used to determine the average heavy nucleus mass fraction, $X_A$. In the current implementation, the three chemical potential constraints combined with charge and baryon number conservation are solved self consistently to determine the matter composition. This leads to a 20\% increase in the mass fraction of heavy nuclei when compared to the original approximation scheme~\cite{Bowers82,Bowers}.

\subsection{Baryonic Matter Above Saturation Density}
Above nuclear matter density, the baryons are treated as a continuous fluid. In this regime, the free energy per nucleon is given in the form 
\begin{equation}
\label{eqn:free_energy}
f = f_1\left(n,Y_p\right) + f_2\left(n, T\right) + 8.79 \ \text{MeV} 	~,
\end{equation}
where the addition of 8.79 MeV sets the zero for the free energy to be the ground state of~$^{56}$Fe.
For an arbitrary proton fraction $Y_p$ and number density $n$ the zero-temperature contribution to the free energy per nucleon is written as the sum of an isospin symmetric term and the symmetry energy:
\begin{equation}
f_1\left(n,Y_p\right) =  \frac{E}{A}\left(n,Y_p=0.5\right) + S\left(n,Y_p\right)		~.
\end{equation}
Expanding $S(n,Y_p)$ in terms of $\left(1-2Y_p\right)$, and keeping only the leading contribution, the symmetry energy can be written as
\begin{equation}
S\left(n,Y_p\right) = \left(1-2Y_p\right)^2S_0\left(n\right)  	~,
\end{equation}
where $S_0$ can be identified as the symmetry energy.

Above saturation density we include both 2-body ($v_{ij}^{(2)}$) and 3-body ($v_{ijk}^{(3)}$) interactions in the many-nucleon system.  The Hamiltonian of this system is thus given by
\begin{equation}
\hat{H} =  \sum_i\hat{t}_i + \sum_{i<j}v_{ij}^{(2)} + \sum_{i<j<k}v_{ijk}^{(3)}	~,
\label{eqn:Hamiltonian}
\end{equation}
where $\hat{t}_i$ is the one body contribution while $v_{ij}$ and $v_{ijk}$ are the 2 and 3-body interactions, respectively.
In the density functional approach one can parameterize these interactions to describe the ground-state properties of finite nuclei and nuclear matter~\cite{Brink, Moszkowski, Skyrme}.  The microscopic interactions, such as meson exchange, are embedded in the parameters of the density dependent forces.  

Among the most widely used interactions are those of the Skyrme type forces.  In this formulation the two-body potential is given in the form~\cite{Vautherin}:
\begin{widetext}
\begin{align}
\begin{split}
v_{12}^{(2)} &= t_0\left(1+x_0\hat{P_s}\right)\delta\left({\bf r_1 - r_2}\right) 
		+ \frac{1}{2}t_1\left(\delta\left({\bf r_1 - r_2}\right)\hat{k}^2+\hat{k}^{'2}\delta\left({\bf r_1 - r_2}\right)\right)  \\
	&\qquad
	+ t_2{\bf \hat{k}^2}\cdot\delta\left({\bf r_1 - r_2}\right){\bf \hat{k}} + iW_0\left({\bf \hat{\sigma}_1}
		+{\bf \hat{\sigma}_2}\right)\cdot{\bf \hat{k}'} \times \delta\left({\bf r_1 - r_2}\right){\bf \hat{k}}	~,
\end{split}
\label{eqn:2body}
\end{align}
\end{widetext}
where $\hat{P_s}$ is the spin exchange operator, ${\bf r_1}$ and ${\bf r_2}$ are the position vectors in the two-body potential, $x_0$ is the coefficient for the isospin exchange operator, $\hat{{\bf k}}$ and $\hat{{\bf k'}}$ are the momentum and conjugate momentum operators, and $W_0$ is the coefficient of the two-body spin orbit interaction. 

We will discuss the Skyrme coefficients \mbox{$t_0, t_1, t_2, t_3,\text{ and }\sigma$} in the following sections.  For this Skyrme potential the high density behavior can be dominated by a 3-body repulsive interaction. This term is taken to be a zero range force of the form $v_{123} = t_3\delta \left({\bf r_1 - r_2}\right)\delta \left({\bf r_2 - r_3}\right)$.  If the assumption is made that the medium is spin-saturated, which is valid for neutron star matter and nuclei~\cite{Ring}, the three-body term is equivalent to a density dependent two-body interaction given by~\cite{Vautherin}
\begin{equation}
v_{12}^{(3)} = \frac{1}{6}t_3\left(1+\hat{P_s}\right)\delta \left({\bf r_1 - r_2}\right) n \left(\frac{{\bf r_1+r_2}}{2}\right).
\end{equation}
In the present formulation we generalize this potential to a modified Skyrme interaction that replaces the linear dependence on the density by a power-law index $\sigma$. This modified Skyrme potential can then be written as~\cite{Mansour}
\begin{equation}
v_{12}^{(3)'} = \frac{1}{6}t_3\left(1+\hat{P_{s}}\right)\delta \left({\bf r_1 - r_2}\right) n^\sigma \left(\frac{{\bf r_1+r_2}}{2}\right).
\label{eqn:3body}
\end{equation}
This modification has been introduced~\cite{Brink} to increase the compressibility of nuclear matter at high densities. A value of $\sigma$ = 1/3 is a common choice~\cite{Kohler, Krivine}.  However, in the present approach we choose to treat $\sigma$ as a free parameter to be determined by constraining the third derivative of the energy per particle (e.g. skewness coefficient) from observed neutron-star properties~\cite{Demorest}.

The main advantage of the Skyrme density functional is that the variables that characterize nuclear matter can be expressed as analytic functions. 
The isospin symmetric contribution is described by a Skyrme density functional with a modified three-body interaction term $\sigma$. 
We use $T_F$ to denote the kinetic energy of a particle at the Fermi surface
\begin{equation}
T_F = \frac{\hbar^2}{2m}\left(\frac{3\pi^2}{2}\right)^{2/3}n^{2/3}  ~.
\end{equation} 
Then, calculating the expectation value of the Hamiltonian [Eq.~(\ref{eqn:Hamiltonian})] in a Slater determinant and setting N=Z, the energy per nucleon for symmetric nuclear matter can be derived~\cite{Vautherin}, as given in Eq.~(\ref{eqn:skyrme}).

All quantities and coefficients for symmetric nuclear matter are obtained from Eq.~(\ref{eqn:skyrme}).
The pressure, is deduced from \mbox{$P = n^2 \frac{\partial}{\partial n} \left( \frac{E}{A} \right)$} and is given in Eq.~(\ref{eqn:skyrme_press}).
Eq.~(\ref{eqn:skyrme_comp}) gives the volume compressibility of symmetric nuclear matter. This is calculated as the derivative of the pressure with respect to number density: 
	\mbox{$K = 9 \left(\frac{\partial P}{\partial n}\right) $} \mbox{$= 18\frac{P}{n}+9n^2 \frac{\partial^2}{\partial n^2} \left( \frac{E}{A}\right)$}.  Finally, the skewness coefficient, \mbox{$Q_0 = 27n^3\frac{\partial^3}{\partial n^3} \left(\frac{E}{A} \right) $}, is deduced from the third derivative of the free energy per nucleon~\cite{Dutra} and is given in Eq.~(\ref{eqn:skyrme_skew}).

\begin{widetext}
	\begin{align}
		\label{eqn:skyrme}
		\frac{E}{A} &= \frac{3}{5}T_F + \frac{3}{8}t_0n + \frac{1}{16}t_3n^{\sigma+1}  
			+ \frac{3}{40}\left(3t_1+5t_2\right)\left(\frac{3 \pi^2}{2}\right)^{2/3}n^{5/3}    \\
		\label{eqn:skyrme_press}
		P 
			&= \frac{2}{5}T_Fn + \frac{3}{8}t_0n^2 + \frac{1}{16}t_3\left(\sigma+1\right)n^{\sigma+2} 
				+ \frac{1}{8}\left(3t_1+5t_2\right)\left(\frac{3\pi^2}{2}\right)^{2/3}n^{8/3}     \\
		\label{eqn:skyrme_comp} 
		K 	
			&= 6T_F + \frac{27}{4}t_0n + \frac{9}{16}t_3(\sigma+1)(\sigma+2)n^{\sigma+1}
			+ 3(3t_1+5t_2)\left(\frac{3\pi^2}{2}\right)^{2/3}n^{5/3}    \\
		\label{eqn:skyrme_skew}
		Q	
			&= \frac{24}{5}T_F 
				+ \frac{27}{16}t_3\sigma\left(\sigma+1\right)\left(\sigma-1\right)n^{\sigma+1} 
				- \frac{3}{4}\left(3t_1+5t_2\right)\left(\frac{3\pi^2}{2}\right)^{2/3}n^{5/3} 
	\end{align}
\end{widetext}

These four equations completely describe the properties of symmetric nuclear matter. Values for the coefficients can be constrained by fixing the density of nuclear saturation, as well as imposing the observational constraint that the maximum mass of a neutron star must exceed $1.97 \pm 0.04 ~M_\odot$~\cite{Demorest}.

We employ a linearly increasing density dependent 
symmetry energy~\cite{Muther, Mayle}.  This behavior is understood to arise from the use of a vector coupled $\rho$-meson in relativistic field theory calculations. The form adopted here for the symmetry coefficient $S_0$ is then
\begin{equation}
S_0(n) = \eta\left[16 + \frac{72}{1+4\eta}\right]  ~,
\label{eqn:sym_energy}
\end{equation}
where $\eta \equiv n/n_0$ is the saturation density parameter with $n_0$ the saturation number density (0.16 fm$^{-3}$)~\cite{BALi}. Therefore, we find the zero-temperature contribution to the free energy to be
\begin{widetext}
\begin{align}
\begin{split}
\label{eqn:total_free}
f_1\left(n,Y_p\right) &= \frac{3}{5}T_F + \frac{3}{8}t_0n + \frac{1}{16}t_3n^{\sigma+1}    
                                   + \frac{3}{40}\left(3t_1+5t_2\right)\left(\frac{3 \pi^2}{2}\right)^{2/3}n^{5/3}  \\
                                   &\qquad \quad
                                   + \eta\left[16 + \frac{72}{1+4\eta}\right]\left(1-2Y_p\right)^2	~.
\end{split}
\end{align}
\end{widetext}

\subsection{Thermal Correction}
For the thermal contribution to the free energy per particle we follow the approach described in Refs.~\cite{WilsonMathews,Mayle}. We assume a degenerate gas 
of the baryonic states reviewed in~\cite{PDGwhite},
as a function of temperature $T$ and baryon density $n$.  Since the zero-temperature contribution to the free energy is already properly taken into account by the Skyrme and symmetry energy contributions, only the thermal portion needs to be added. We also assume that the 
baryonic states
are in chemical equilibrium. The expression for the thermal contribution is written:
\begin{equation}
	f_2(n,T) = \alpha-\alpha_0 + \frac{1}{n}\left(\omega - \omega_0\right) ~,
\end{equation}
where $\alpha$ and $\omega$ are the finite temperature ``chemical potential'' and the grand potential density, respectively. The quantities, $\alpha_0$ and $\omega_0$ are the zero-temperature limits of the ``chemical potential'' and grand potential density and $n$ is the local baryon number density.

$\alpha_0$ is constrained from the number density of baryons and is determined from the relation
\begin{equation}
n = \sum_i\frac{g_i}{2\pi^2}\left(\alpha_0^2 - \widetilde{m}^2_i\right)^{3/2} ~,
\label{eqn:numdensity}
\end{equation}
while the zero-temperature limit of the grand potential density is
\begin{widetext}
\begin{equation}
\omega_{0} = -\sum_i\frac{g_i}{2\pi^2}\left[\alpha_0\sqrt{\alpha_0^2-\widetilde{m}^2_i}\left(\alpha_0^2-\frac{5}{2}\widetilde{m}_i^2\right)+\frac{3}{2}\widetilde{m}_i^4\text{ln}\left(\frac{\alpha_0+\sqrt{\alpha_0^2-\widetilde{m}_i^2}}{\widetilde{m}_i}\right)\right]   ~.
\label{eqn:omega0}
\end{equation}
\end{widetext}
In Eqs (\ref{eqn:numdensity}) and (\ref{eqn:omega0}) $\widetilde{m}_i$ is an effective particle mass deduced from fits to results from relativistic Bruckner Hartree-Fock theory~\cite{Migdal}
\begin{equation}
\label{eqn:eff_mass}
\widetilde{m}_i = \frac{m_i}{1+0.27\eta} ~,
\end{equation}
and the sum over i includes both nucleons and delta particles.

The finite temperature $\alpha$ is then found from baryon number conservation
\begin{equation}
	n = \sum_i\frac{g_i}{2\pi^2}\int_0^{\infty}\left(h(p,\alpha)-h(p,-\alpha)\right)p^2 dp	,
\label{eqn:numberdensity}
\end{equation}
and the finite temperature grand potential density is given as
\begin{equation}
\omega = -\sum_i\frac{g_i}{6\pi^2}\int_0^\infty\frac{p^4}{\epsilon_i}\left(h(p,\alpha) + h(p,-\alpha)\right)dp ,
\label{eqn:omega}
\end{equation}
where $\epsilon_i$ is given by
\begin{equation}
\epsilon_i = \sqrt{p^2+\widetilde{m}_i^2}.
\end{equation}
In Eqs.~(\ref{eqn:numberdensity}) and~(\ref{eqn:omega}), $h(p,\alpha)$ is just the usual Fermi distribution function \begin{equation}
	h(p_i,\alpha_i) = \frac{1}{\exp[(\epsilon_i - \alpha_i)/T] + 1}  	~,
\end{equation}
and $g_i$ is the spin-isospin degeneracy factor.  It should be noted that $\alpha$ and $\alpha_0$ are only used to construct $f_2$ and do not correspond to a real chemical potential. The actual chemical potentials are found from derivatives of the total free energy~[Eq.~\ref{eqn:free_energy}] with respect to density. They are given as
\begin{align}
\label{eqn:mun}
\mu_n &= f + n\left(\frac{\partial{f}}{\partial{n}}\right)_{T,Y_p} - Y_p\left(\frac{\partial{f}}{\partial{Y_p}}\right)_{T,n} \\
\mu_p &= f + n\left(\frac{\partial{f}}{\partial{n}}\right)_{T,Y_p} + Y_n\left(\frac{\partial{f}}{\partial{Y_p}}\right)_{T,n}
\label{eqn:mup}
\end{align}
where $Y_n + Y_p = 1$. 

\subsection{Thermodynamic State Variables}
Once the thermal contribution to the free energy is constructed the thermodynamic quantities can be calculated. Of particular interest are the total internal energy ($E$), the total pressure ($P$), the entropy per baryon ($S$) and the adiabatic index $\Gamma$. The total internal energy is calculated from the free energy:
\begin{equation}
E = f - T\left(\frac{\partial f}{\partial T}\right)_{n,Y_e} + E_{e} + E_{\gamma}    ~,
\end{equation}
where $E_e$ is the electron energy determined by numerically integrating over Fermi-Dirac distributions and $E_{\gamma}$ is the photon energy contribution determined from the usual Stefan-Boltzmann law.

The pressure is calculated from \mbox{$P = n^2\left(\partial f/ \partial n\right)$}. It is important to note that the thermal contribution to the pressure is not the simple form of $\omega_0 - \omega$ as one would expect from the usual application of the thermodynamic potential, but is given by a slightly more complicated form
\begin{equation}
\label{eqn:therm_press}
P_{therm} = \omega_0 - \omega +n\frac{\partial}{\partial n}\left[\omega - \omega_0\right].
\end{equation}
This is due to the fact that the effective mass [cf.~Eq.~(\ref{eqn:eff_mass})] is density dependent. If this dependence were removed we would recover the usual form of the pressure from the thermodynamic potential. 

We thus determine the total pressure from the free energy 
\begin{equation}
\label{eqn:total_press}
P = n^2\left(\frac{\partial f_1}{\partial n}\right) + P_{therm} + P_e + P_{\gamma} ~,
\end{equation}
where again the electron and photon contributions are determined from the pressures calculated previously. The entropy per baryon in units of Boltzmann's constant is given by a simple derivative \mbox{$S = -\left(\partial f/ \partial T\right)$}, and the adiabatic index is given by the usual form \mbox{$\Gamma = \left[\partial \ \text{ln} P/ \partial \ \text{ln} (n)\right]$}.

To investigate the physical quantities relevant for modeling nuclear matter in dense astrophysical environments we determine the coefficients in Eq.~(\ref{eqn:skyrme}) by utilizing known constraints on nuclear matter saturation. Applying the saturation condition \mbox{$P(n=n_0) = n^2 \frac{\partial}{\partial n}\left(\frac{E}{A}\right) |_{n=n_0} = 0$} to \mbox{Eqs.~(\ref{eqn:skyrme}-\ref{eqn:skyrme_skew})}, one gets a system of four equations in terms of the quantities $t_0$, $\left(3t_1+5t_2\right)$, $t_3$, and $\sigma$:
\begin{widetext}
\begin{align}
E_0 &= \frac{3}{5}T_{F_0} + \frac{3}{8}t_0n_0 + \frac{1}{16}t_3n_0^{\sigma+1}+\frac{3}{40}\left(\frac{3\pi^2}{2}\right)^{2/3}\left(3t_1 + 5t_2\right)n_0^{5/3}  \label{eqn:Energy1}\\
P_0 &= 
     0 = \frac{2}{5}T_{F_0} + \frac{3}{8}t_0n_0 + \frac{1}{16}t_3\left(\sigma+1\right)n_0^{\sigma+1}+\frac{1}{8}\left(\frac{3\pi^2}{2}\right)^{2/3}\left(3t_1 + 5t_2\right)n_0^{5/3} \label{eqn:pressure1} \\
K_0 &= -\frac{6}{5}T_{F_0} + \frac{9}{16}t_3\sigma\left(\sigma+1\right)n_0^{\sigma+1} + \frac{3\left(\frac{3\pi^2}{2}\right)^{2/3}}{4}\left(3t_1 + 5t_2\right)n_0^{5/3} \label{eqn:compressibility1} \\
Q_0 &= \frac{24}{5}T_{F_0} + \frac{27}{16}t_3\sigma\left(\sigma+1\right)\left(\sigma-1\right)n_0^{\sigma+1}-\frac{3\left(\frac{3\pi^2}{2}\right)^{2/3}}{4}\left(3t_1 + 5t_2\right)n_0^{5/3} \label{eqn:Skewness1}
\end{align}
\end{widetext}
Solving \mbox{Eqs.~(\ref{eqn:Energy1}) - (\ref{eqn:Skewness1})} for $\sigma$ then yields
\begin{equation}
\sigma = \frac{\frac{9}{5}T_{F_0}-2K_0 + Q_0 - 45E_0}{3K_0 + 45E_0 - \frac{27}{5}T_{F_{0}}}  ~.
\end{equation}

The usual approach is to choose a set of data, e.g. resonances, nuclear masses, charge radii, etc., to find a best set of parameters for a Skyrme model. From these Skyrme coefficients, then, the quantities in \mbox{Eqs.~(\ref{eqn:Energy1}) - (\ref{eqn:Skewness1})} are deduced, i.e. $n_0, T_{F_0}, E_0, K_0, Q_0$. For our purposes, however, we choose a more empirical approach. That is, we adopt inferred values of $n_0, E_0, K_0, Q_0$ from the literature and use these to determine the Skyrme model parameters. We also demand that these parameters allow neutron star \mbox{masses $\ge 1.97 \pm 0.04 ~M_\odot$}.  In this way our EoS relates directly to inferred properties of nuclear matter in the literature and our EoS is easily adaptable to improved experimental and theoretical determinations. 

\begin{table} [t]
\begin{tabular} {| c  c  c  | }
\hline
 Parameter            & value               &  ref. \\
\hline
$n_0$  &            0.16 $\pm$ 0.01 fm$^{-3}$     &             \cite{BALi}     \\ 
$E_0$  &            -16 $\pm$ 1 MeV                     &             \cite{BALi}     \\ 
$K_0$  &            240 $\pm$ 10 MeV                 &             \cite{Colo}           \\
$Q_0$  &             -390 $\pm$ 90 MeV                &         this work         \\
\hline
\end{tabular} 
\caption{Adopted constraints on properties of nuclear matter.}
\label{tble:Skyrme}
\end{table}

The saturation density \mbox{$n_0 \approx 0.16 ~\text{fm}^{-3}$} and the binding energy per nucleon \mbox{$E_0 = -16$ MeV} are reasonably well established~\cite{BALi}. The determination of the compressibility parameter from experimental data on the giant monopole resonance on finite nuclei, however, has been a long standing conundrum. On the one hand, the compressibility of nuclear matter can be determined~\cite{Farine} by fitting measured breathing-mode energies, using generalized Skyrme-type forces that include a density and momentum dependent term. Acceptable fits are in the range of \mbox{$K_0 = 215 \pm 15$ MeV}~\cite{Farine}. On the other hand a value of \mbox{$K_0 = 231 \pm 5$ MeV}, has been found~\cite{Youngblood} by using measured E0 distributions in $^{40}$Ca, $^{90}$Zr, $^{116}$Sn, $^{144}$Sm and $^{208}$Pb based upon the calculations of~\cite{Blaizot}. Building a new class of Skyrme forces in Ref.~\cite{Colo} a value of \mbox{$K_0 = 240 \pm 10$ MeV} was found from these data. 

Given the unresolved discrepancy between GMR data and models~\cite{Dutra}, there is currently a fairly large uncertainty in $K_0$. However, for our purposes we adopt the median value and uncertainty from Ref.~\cite{Colo}, i.e. \mbox{$K_0$ = 240 $\pm$ 10 MeV} as this is most appropriate for the Skyrme force approach employed here. Solving \mbox{Eqs.~(\ref{eqn:Energy1}) - (\ref{eqn:Skewness1})} self consistently, we therefore determine the best range for the nuclear compressibility consistent with the results of~\cite{Colo}.  

\begin{figure}[t]
  \centering
  \includegraphics[width=0.5\textwidth]{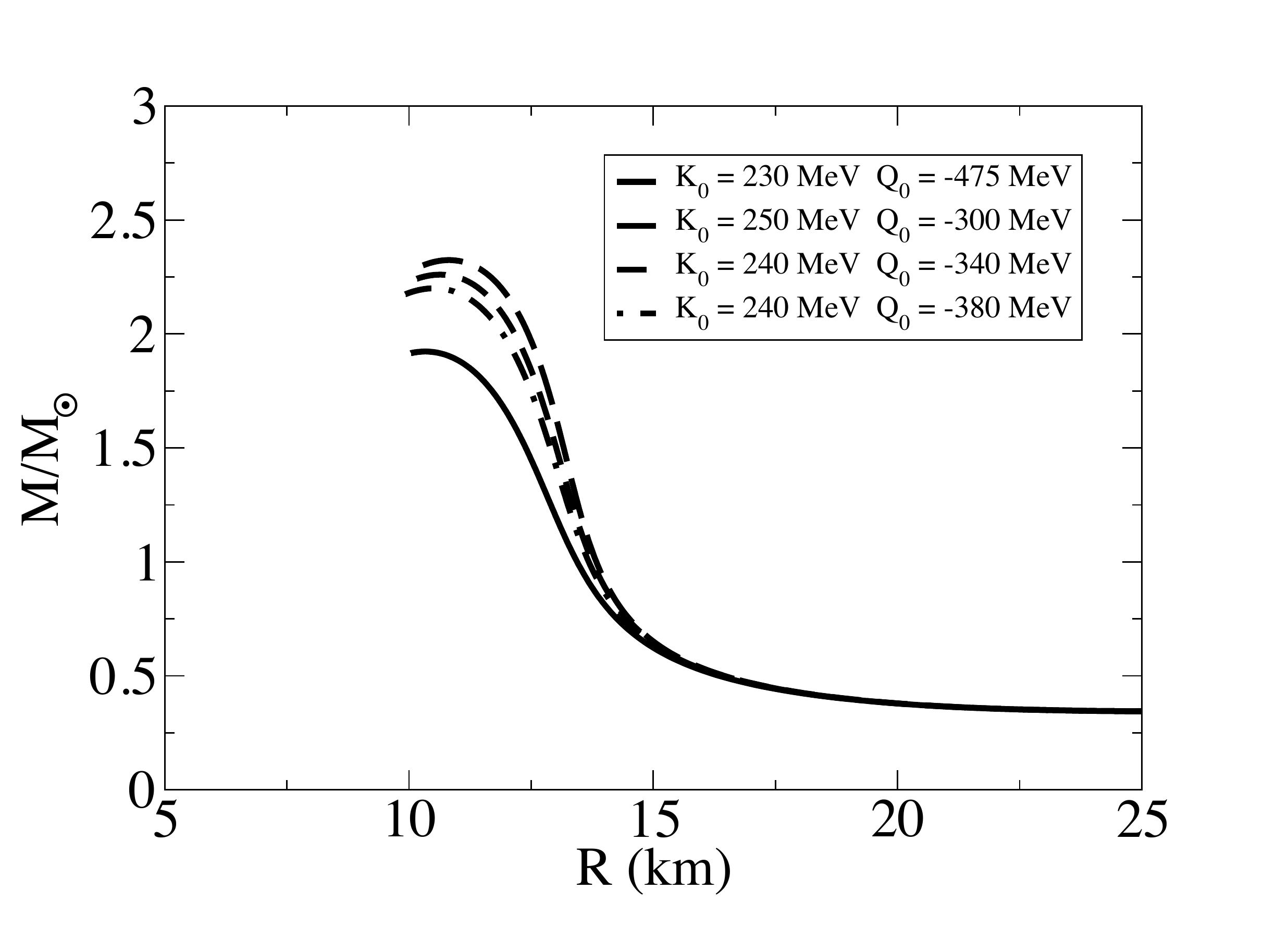}
  \caption{Mass vs. radius diagram constraining the skewness coefficient $Q_0$. The lower bound of $Q_0 >$ -475 MeV is constrained by the \mbox{$M_{max} \ge 1.97 \pm 0.04 ~M_\odot$}~\cite{Demorest} observational limit.}
    \label{fig:MaxMass_Q0}
\end{figure}

There is even more uncertainty in the skewness parameter $Q_0$. In Ref.~\cite{Farine}, breathing mode data were used to find a weak inverse correlation between $Q_0$ and $K_0$. In that work however, they could only deduce a very broad range for \mbox{$Q_0 = -700 \pm 500$ MeV}.  We find an upper bound on the skewness coefficient by using the stiffest compressibility in our range (\mbox{$K_0 = 250$ MeV}) in \mbox{Eqs.~(\ref{eqn:Energy1}) - (\ref{eqn:Skewness1})}.  This gives us an
upper bound on the skewness coefficient of \mbox{$Q_0 = -300$ MeV}.  
The lower bound on $Q_0$ is found similarly, using the smallest compressibility in our range (\mbox{$K_0 = 230$ MeV}).  It is further constrained that the 
maximum mass of the neutron star is above the \mbox{$1.97 \pm 0.04 ~M_\odot$} observation.  
From Fig.~\ref{fig:MaxMass_Q0}, the lowest skewness coefficient that meets this constraint is \mbox{$Q_0 = -387 \pm 87$ MeV}.  This is also consistent within the range given in Ref.~\cite{Colo}. The fiducial NDL~EoS is constructed using the median values of the constraints given from \mbox{Table~\ref{tble:Skyrme}}. 

The Skyrme coefficients for the fiducial NDL~EoS are then found to be 
\begin{gather*}
	t_0 =  -1718 \text{ MeV fm}^{3}                               \\
	\left(3t_1+5t_2\right) =  -102 \text{ MeV fm}^5      \\
	t_3 =  13226 \text{ MeV fm}^{3\sigma+3}              \\
	\sigma = 0.369	~.
\end{gather*}	
Note that our deduced value of the 3-body index, $\sigma$, is close to the commonly employed value of 1/3~\cite{Kohler, Krivine}.

The density dependence of the symmetry energy beyond saturation is highly uncertain. For many Skyrme models the symmetry energy either saturates at high densities, or in the worst case becomes negative. This results in a negative pressure deep inside the neutron star core. For this work, we choose a fairly stiff symmetry energy, Eq.~(\ref{eqn:sym_energy}). That is, we implemented a linearly increasing function of density to remove the issues inherent to many Skyrme parameter sets.  The symmetry energy at saturation is known~\cite{Lattimer2012} to lie within the range \mbox{$S_0 = 26 - 34$ MeV} 
and is determined by the difference between the energy per particle for pure neutron matter and that of symmetric matter at \mbox{$T = 0$ MeV}. For all relevant parameter sets chosen the NDL~EoS symmetry energy at saturation is found to be \mbox{$S_0 = 30.4$ MeV} (see Fig.~\ref{fig:sym_energy}).

\begin{figure}[t!]
  \centering
  \includegraphics[width=0.5\textwidth]{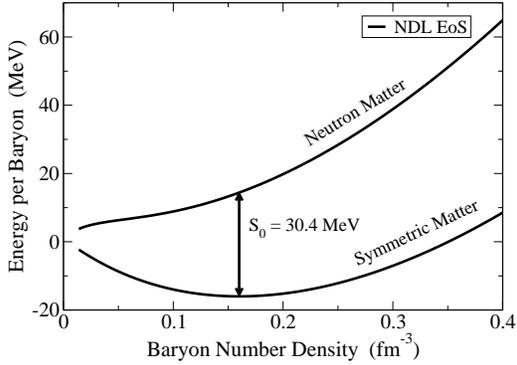}
  \caption{The energy per baryon for pure neutron matter and symmetric nuclear matter. The symmetry energy is calculated as the difference between these functions at saturation density. For the NDL~EoS a symmetry energy of $S_0$ = 30.4 MeV was found.}
 \label{fig:sym_energy}
\end{figure}

\section{Pions in the nuclear environment}
The impact on the hadronic EoS from the lightest mesons (i.e. the pions) has been constrained~\cite{McAbee} from a comparison between relativistic heavy-ion collisions and one-fluid nuclear collisions.  The formation and evolution of the pions was computed in the context of Landau-Migdal theory~\cite{Migdal} to determine the pion effective energy and momentum. In this approach the pion energy is given by a dispersion relation~\cite{Migdal}
\begin{equation}
\epsilon_\pi^2 = p_\pi^2 + \tilde{m}_\pi^{2}    ~,
\label{eqn:dispersion}
\end{equation}
where $\tilde{m}_\pi$ is the pion ``effective mass'' defined to be
\begin{equation}
\tilde{m}_\pi = m_\pi\sqrt{1+\Pi\left(\epsilon_\pi,p_\pi,n\right)}   ~.
\label{eqn:pionmass}
\end{equation}

Following~\cite{Mayle} and~\cite{Friedman} the polarization parameter $\Pi$ can be written, 
\begin{equation}
\Pi\left(\epsilon_\pi,p_\pi,n\right) = \frac{p_\pi^2 \Lambda^2\left(p_\pi\right)\chi \left(\epsilon_\pi,p_\pi,n\right)}{m_\pi^2-g'm_\pi^2\Lambda^2\left(p_\pi\right)\chi\left(\epsilon_\pi,p_\pi,n\right)}   ~.
\label{eqn:propagator}
\end{equation}
where the denominator is the Ericson-Ericson-Lorentz-Lorenz correction~\cite{Ericson}. The quantity \mbox{$\Lambda \equiv \text{exp}(-p_\pi^2/b^2)$} with \mbox{$b = 7m_\pi$}, is a cutoff that ensures that the dispersion relation~[Eq.~(\ref{eqn:dispersion})] asymptotically approaches the high momentum limit, 
\begin{equation}
\epsilon_{\infty}\equiv\epsilon\left(p_\pi\rightarrow\infty,n\right) = \sqrt{m_\Delta^2+p_\pi^2}-m_N ~.
\end{equation}

Following~\cite{Ericson} we take the polarizability to be
\begin{equation}
\chi\left(\epsilon_\pi,p_\pi,n\right) = -\frac{4a\epsilon_{\infty}n}{\epsilon_\infty^2-\epsilon_\pi^2} ~,
\end{equation}
where \mbox{$a = 1.13/m_\pi^2$}.  This form for the polarizability ensures that the effective pion mass is always less than or equal to the vacuum rest mass $m_\pi$.

A key quantity in the above expressions is the Landau parameter $g'$. This is an effective nucleon-nucleon coupling strength.  To ensure consistency with observed Gamow-Teller transition energies a constant value of \mbox{$g' = 0.6$} was used in~\cite{Friedman}. However, in~\cite{McAbee}, Monte-Carlo techniques were used to statistically average a momentum dependent $g'$ with particle distribution functions. It was found that $g'$ varies linearly with density and is approximately given by
\begin{equation}
	g' = g_1 + g_2\eta   ~.
\end{equation}
 A value of \mbox{$g_1 = 0.5$} was chosen to be consistent with known Gamow-Teller transitions. A value for $g_2$ was then obtained~\cite{McAbee} by optimizing fits to a range of pion multiplicity measurements obtained at the Bevlac~\cite{Harris}. These data were best fit for a value of \mbox{$g_2 = 0.06$}. 
 
 The pions are assumed to be in chemical equilibrium with the surrounding nuclear matter. We consider the pion-nucleon reactions:
\begin{equation}
p\leftrightarrow n + \pi^+  ~, \quad   n\leftrightarrow p + \pi^-     ~.
\end{equation}
This leads to the following relations among the chemical potentials for neutrons, protons, and pions
\begin{equation}
\mu_p = \mu_n + \mu_{\pi^+}  ~, \quad   \mu_n = \mu_p + \mu_{\pi^-}  ~.
\end{equation}
These equilibrium conditions let us express the pion chemical potentials in terms of the neutron and proton chemical potentials: \mbox{$\hat{\mu} \equiv \mu_n - \mu_p$} $= \mu_{\pi^-} = -\mu_{\pi^+}$.
Using the definitions of  $\mu_n$ and $\mu_p$ from \mbox{Eqs.~(\ref{eqn:mun}) - (\ref{eqn:mup})}, 
the expressions for the pion chemical potentials are found to be
\begin{equation}
\label{eqn:mupi}
\mu_{\pi^-} = -\mu_{\pi^+} = \frac{4S\left(n,Y_p\right)}{\left(1-2Y_p\right)}.
\end{equation}
Where $S(n,Y_p)$ is the nuclear symmetry energy from Eq.~(\ref{eqn:sym_energy}).

For a given temperature ($T$) and number density ($n$) the pion number densities are given by the standard Bose-Einstein integrals
\begin{equation}
\label{eqn:pin}
n_i = \int_0^\infty\frac{p^2}{2\pi^2}\frac{dp}{e^{(\epsilon_\pi - \mu_i)/T}-1}   ~,
\end{equation}
where $i$ sums over $\{ \pi^+, \pi^-, \pi^0 \}$, and $\epsilon_\pi$ is given by Eq.~(\ref{eqn:dispersion}). Note that the $\pi^0$ chemical potential is taken to be zero, since these particles can be created or destroyed without charge constraint.  

The charge fraction per baryon for the charged pions is defined as \mbox{$Y_{\pi^-} = n_{\pi^-}/n$}.  From Eq.~(\ref{eqn:mupi})  we can calculate the pion number densities from the pion chemical potentials. Then, electric charge conservation gives,
\begin{equation}
\label{eqn:Yp}
Y_e = Y_p  - Y_{\pi^-} + Y_{\pi^+}   ~.
\end{equation}
Thus, we can solve Eq.~(\ref{eqn:Yp}) for the unknown quantity $Y_p$.

Once $Y_p$ is determined, the pionic energy densities and partial pressures can be calculated from
\begin{equation}
E_i = \int^\infty_0\frac{p^2}{2\pi^2}\frac{\epsilon_\pi}{\text{exp}\left[\left(\epsilon_\pi-\mu_i\right)/T\right]-1}  ~,
\end{equation}
and
\begin{equation}
P_i = \int^\infty_0\frac{p^2}{2\pi^2}\frac{(1/3)p(\partial{\epsilon_\pi}/\partial{p})}{\text{exp}\left[\left(\epsilon_\pi-\mu_i\right)/T\right]-1}   ~.
\end{equation}
Note that in the high temperature, low-density regime we add all baryonic and mesonic resonances.  In this limit the pionic mass approaches the bare pion mass.  Hence, we also trial all mesonic and baryonic states using bar masses.

\section{QCD Phase Transition}
It is generally expected~\cite{McLerran} that for sufficiently high densities and/or temperature, a transition from hadronic matter to quark-gluon plasma (QGP) can occur.  Recent progress~\cite{Kronfeld} in lattice gauge theory (LGT) has shed new light on the transition to a  QGP in the low baryochemical potential, high-temperature limit. It is now believed that at high temperature and low density a deconfinement and chiral symmetry restoration occur simultaneously at the crossover boundary. In particular, at low density and high temperature, it has been found~\cite{Kronfeld} that the order parameters for deconfinement and chiral symmetry restoration changes abruptly for temperatures of \mbox{$T = 145 - 170$ MeV}~\cite{Borsanyi, Bazavov1}. However, neither order parameter exhibits the characteristic change expected from a 1st order phase transition. An analysis of many~\cite{Aoki,Bazavov2} thermodynamic observables confirms that the transition from a hadron phase to a high temperature QGP is a smooth crossover. 

At low density the hadron phase can be approximated as a pion-nucleon gas, while the QGP phase can be approximated as a non-interacting relativistic gas of quarks and gluons~\cite{Fuller}. Equating the pressures in the hadronic and QGP phases, the critical temperature $T_c$ for the low density transition can be approximated~\cite{Fuller} as:
\begin{equation}
T_c \approx \left(g_q - g_h\right)^{-1/4}\left(\frac{90}{\pi^2}\right)^{1/4}B^{1/4}  ~.
\label{eqn:Tc}
\end{equation}
Where the statistical weight $g_q$ for a low-density high-temperature QGP gas with three relativistic quarks is \mbox{$g_q \approx 51.25$}, while \mbox{$g_h \approx 17.25$} was found for the hadronic phase by summing over all known meson data. 

Adopting the lattice gauge theory results~\cite{Kronfeld} that  \mbox{$145 \lesssim T_c \lesssim 170$ MeV}, then implies~\cite{Fuller} that a reasonable range for the QCD vacuum energy is \mbox{$165 \lesssim B^{1/4} \lesssim 240$ MeV}.  
This provides an initial range for the QCD vacuum energy.  We will further constrain this parameter by requiring that the maximum mass of a neutron star exceed \mbox{$1.97 \pm 0.04 ~M_\odot$}~\cite{Demorest}.

Another parameter that impacts the thermodynamic properties of the system is the strong coupling constant $\alpha_s$.  For this manuscript we adopt a value of $\alpha_s = 0.33$ as this is a representative value for the energy regime under consideration~\cite{PDGwhite}.



A transition to a QGP phase during the collapse can have a significant impact on the dynamics and evolution of the nascent proto-neutron star. In~\cite{Gentile} it was first shown that a first order phase transition to a deconfined QGP phase resulted in the formation of two distinct but quickly coalescing shock waves.  More recently, it has been shown~\cite{Fischer2010} that if the transition is first order, but global conservation laws are invoked, then the two shock waves can be time separated by as much \mbox{as $\sim 150$ ms}. Neutrino light curves showing such temporally separated spikes might even be resolvable in modern terrestrial neutrino detectors~\cite{Fischer}. 

The observation of a \mbox{$1.97 \pm 0.04 ~M_\odot$} neutron star, however, 
constrains the possibility
of a first order phase transition to a quark gluon plasma taking place inside the interiors of stable cold neutron stars~\cite{Demorest}. Nevertheless, for initial stellar masses \mbox{beyond $\gtrsim 20 ~M_\odot$} every phase of matter must be traversed during the formation of stellar mass black holes.  Hence at the very least, this transition to QGP may have an impact~\cite{Nakazato} on the neutrino signals during black hole formation as well as its possible impact on core-collapse supernovae.

\subsection{The Quark Model}
For the description of quark matter we utilize a bag model with 2-loop corrections, and construct the EoS from a phase-space integral representation over scattering amplitudes.  We allow for the possibility of a coexistence mixed phase in a first order transition, or a simple direct cross over transition. In the hadronic phase the thermodynamic state variables, are calculated from the Helmholtz free energy $F(T,V,N)$ as described in the previous sections.  However, it is convenient to compute the QGP in terms of  the grand potential, $\Omega(T,V,\mu)$.  Both descriptions are equivalent and are related  by a Legendre transform: 
	$\Omega = F - \sum_i \mu_iN_i$.

The grand potential for  the quark-gluon plasma takes the form:
\begin{equation}
\Omega = \sum_i(\Omega_{q0}^i + \Omega_{q2}^i) +  \Omega_{g0} + \Omega_{g2} + B V .
\label{eq:GrandPotential}
\end{equation}
Where $q_0$ and $g_0$ denote the $0^\text{th}$-order bag model thermodynamic potentials for quarks and gluons, respectively, while $q_2$ and $g_2$ denote the 2-loop corrections. In most calculations sufficient accuracy is obtained by using fixed current algebra masses (e.g. \mbox{$m_u \sim m_d \sim 0$ GeV}, \mbox{$m_s \sim 0.1 - 0.3$ GeV}). For this work we chose the strange quark mass to be $m_s$ = 150 MeV and a bag constant $B^{1/4} = 165 - 240$ MeV.  The quark contribution to the thermodynamic potential is  given~\cite{McLerran} in terms of a sum of the ideal gas contribution plus a two loop correction from phase-space integrals over Feynman amplitudes~\cite{Kapusta}:
\begin{widetext}
	\begin{align}
		\label{eqn:idealgas} 
		\Omega_{q0}^{i} =& -2N_cT \int_0^{\infty} \frac{d^3p}{\left(2\pi^2\right)} 
			\left[\text{ln}\left(1+e^{-\beta\left(E_i-\mu_i\right)}\right) 
				+ \text{ln}\left(1+e^{-\beta\left(E_i+\mu_i\right)}\right)\right]    \\
		\label{eqn:twoloop}
		\Omega_{q2}^i =& \alpha_s \pi \left(N_c^2-1\right) \Bigg[
			\frac{1}{3}\int_0^\infty \frac{d^3p}{\left(2\pi^2\right)}\frac{N_i(p)}{E_i(p)}
				+ \int_0^\infty \frac{d^3p}{\left(2\pi^2\right)} \frac{d^3p'}{\left(2\pi^2\right)}
					\frac{1}{E_i(p)E_i(p')}\left[N_i(p)N_i(p')+2\right] 	\nonumber \\
			  &\qquad \qquad \quad \times \left[
			  \frac{N_i^+(p)N_i^+(p')+N_i^-(p)N_i^-(p')}{\left(E_i(p)-E_i(p')\right)^2-\left({\bf p-p'}\right)^2} 
			  + \frac{N_i^+(p)N_i^+(p')+N_i^-(p)N_i^-(p')}{\left(E_i(p)-E_i(p')\right)^2-\left({\bf p-p'}
			  \right)^2}\right]\Bigg] ,
\end{align}
\end{widetext}
where the $N_i^{\pm}$ denote Fermi-Dirac distributions:
\begin{equation}
N_i^{\pm}(p) = \frac{1}{e^{\beta\left(E_i(p)\mp\mu_i\right)} + 1}   ~.
\end{equation}

The one- and two-loop gluon and ghost contributions to the thermodynamic potentials can be evaluated in a similar fashion to that of the quarks.
\begin{align}
\Omega_{g0} =& 2\left(N_c^2-1\right)T\int_0^\infty\frac{d^3p}{\left(2\pi^2\right)}\text{ln}\left(1-e^{-\beta\lvert p \rvert}\right) \nonumber \\
                         =& -\frac{\pi}{45}\left(N_c^2-1\right)T^4   ~. 
\end{align}
\begin{equation}
\Omega_{g2} = \frac{\pi}{36}\alpha_sN_c\left(N_c^2-1\right)T^4.
\end{equation}

For the massless quarks, \mbox{Eqs.~(\ref{eqn:idealgas}-\ref{eqn:twoloop})} are easily evaluated to give
\begin{align}
\Omega_{q0}^i =& -\frac{N_c}{6}\left(\frac{7\pi^2}{30}T^4 + \mu_i^2T^2 +\frac{\mu_i^4}{2\pi^2}\right) \\
\Omega_{q2}^i =& \frac{(N_c^2-1)\alpha_s}{8\pi}\left(\frac{5\pi^2}{18}T^4 + \mu_i^2T^2 +\frac{\mu_i^4}{2\pi^2}\right)  .
\end{align}
For the massive strange quark Eq.~(\ref{eqn:idealgas}) can be easily integrated. Eq.~(\ref{eqn:twoloop}), however, cannot be integrated numerically, due to the divergences inherent in it. We therefore, approximate~\cite{McLerran} the two loop strange quark contribution with the zero mass limit. This may over estimate the contribution due to a finite strong coupling constant, but given that the quark mass is relatively small compared to its chemical potential, this is a reasonable approximation.

\subsection{Conservation Constraints}
The neutronized matter deep inside the core of a collapsing star consists of a multicomponent system constrained by the conditions of both charge and baryon number conservation.   The pressure varies as a function of density for a first order transition producing a mixed phase of material. In fact, all thermodynamic quantities vary in proportion to the volume fraction~[\mbox{$\chi \equiv V^Q/(V^Q+V^H)$}] throughout the mixed phase regime.

For the description of a first order phase transition we utilize a Gibbs construction. In this case the two phases are in equilibrium when the chemical potentials, temperatures and the pressures are equal. For the description of the phase transition from hadrons to quarks this construction can be written
\begin{eqnarray}
\mu_p &=& 2\mu_u + \mu_d \\
\mu_n &=& 2\mu_d + \mu_u \\
\mu_d &=& \mu_s \\
T_H &=& T_Q   \\
P^H\left(T,Y_e, \{\mu_i^H\}\right) &=& P^Q\left(T,Y_e,\{\mu_i^Q\}\right)   ,
\end{eqnarray}
where \{$\mu_i^H$\} $\equiv$ \{$\mu_n, \mu_p, \mu_e,\mu_\nu$\} and \{$\mu_i^Q$\} $\equiv$ \{$\mu_u,\mu_d,\mu_s,\mu_e,\mu_\nu$\}.  

The Gibbs construction ensures that a uniform background of photons and leptons exists within the differing phases. Therefore, the contribution from the photon, neutrino, electron, and other lepton pressures cancel out in phase equilibrium. Also from this, we find that the two conserved quantities vary linearly in proportion to the degree of completion of the phase transition i.e.
\begin{align}
n_B Y_e =&  \left(1-\chi\right)n_B^HY_c^H + \chi n_B^QY_c^Q \\
n_B =& \left(1-\chi\right)n_B^H + \chi n_B^Q ,
\end{align}
where we have defined \mbox{$Y_c^H = Y_p + Y_{\pi^+} - Y_{\pi^-}$} and \mbox{$n_B^Q Y_c^Q = 1/3~(2n_u - n_d - n_s)$}.  The internal energy and entropy densities likewise vary in proportion to the degree of phase transition completion
\begin{align}
\epsilon =& \left(1-\chi\right)\epsilon^H + \chi \epsilon^Q \\
             s =&  \left(1-\chi\right)s^H + \chi s^Q    .
\end{align}

\section{Results and Comparisons}
In this section we compare properties of the new NDL~EoS with the two most commonly employed equations of state used in astrophysical collapse simulations as well as the original EoS of Bowers \& Wilson.
\begin{figure}[h!]
  \centering
  \includegraphics[width=0.5\textwidth]{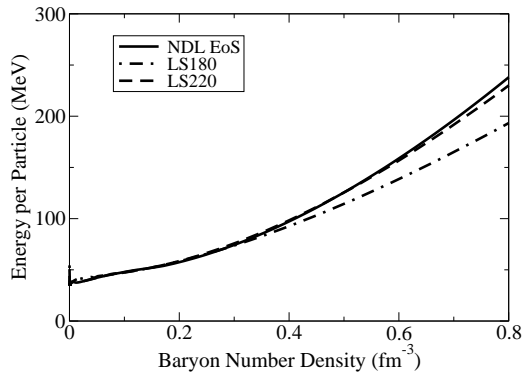}
  \caption{The energy per particle as a function of density comparing the Lattimer~\&~Swesty EoS with compressibilities $K_0$ = 180 MeV and $K_0$ = 220 MeV with the fiducial NDL~EoS.}
 \label{fig:had_energy}
\end{figure}
Fig.~\ref{fig:had_energy} shows the total internal energy as a function of local proper baryon density. We compare the Lattimer \& Swesty EoS~\cite{LS91} at two differing compressibilities (\mbox{$K_0 = 180$ MeV} and \mbox{$K_0 = 220$ MeV}) with the fiducial NDL~EoS 
at a fixed electron fraction and temperature of $Y_e = 0.3$ and $T = 10$~MeV.  A steep rise in the energy per baryon at high densities occurs for larger values of the compressibility as expected.

Similarly, Fig.~\ref{fig:had_pressure} depicts the pressure vs. density for the Shen EoS~\cite{Shen2011}, the Lattimer~\&~Swesty EoS~\cite{LS91} and the NDL~EoS. The Shen EoS consistently leads to higher pressure.  This results from the use of the TM1 parameter set that contains relatively high values for both the symmetry energy at saturation and the nuclear compressibility typical of RMF approaches.

\begin{figure}[h!]
  \centering
  \includegraphics[width=0.5\textwidth]{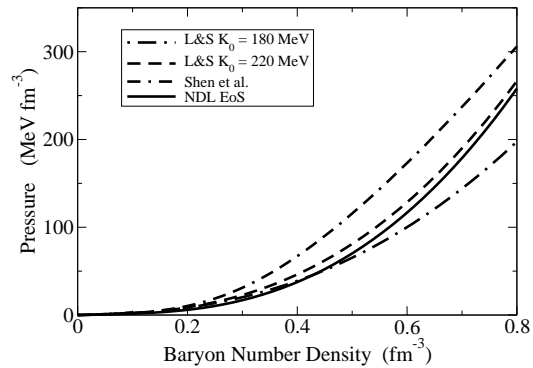}
  \caption{The pressure as a function of density comparing two EoSs from Lattimer \& Swesty~\cite{LS91}, the Shen EoS~\cite{Shen2011} and the fiducial NDL~EoS of the present work.}
    \label{fig:had_pressure}
\end{figure}

\subsection{Pion Effects on the EoS}
The solution to the pion dispersion relation, Eq.~(\ref{eqn:dispersion}), does not produce conditions within the supernova core to generate a pion condensate. The pions considered here are thermal pionic excitations calculated from the pion propagator in Eq.~(\ref{eqn:propagator}). In very hot and dense nuclear matter the number density of pionic excitations is greatly enhanced by the $\pi N \Delta$ coupling~\cite{Friedman}. Hence, it becomes energetically favorable to form pions in the nuclear fluid when the chemical balance shifts from electrons to negative pions. This allows the charge states to equilibrate with these newly formed bosons.  Since the pions are assumed to be in chemical equilibrium with the surrounding nuclear fluid, this has a profound effect on the proton fraction within the medium, particularly for low electron fractions, (see Fig.~\ref{fig:pion_prot_frac}). 

The pion charge fraction as a function of baryon number density is shown in Fig.~\ref{fig:pion_charge_frac}. For a low fixed $Y_e$ the charge fraction of negative pions can actually become greater than the electron fraction, and the negative pions essentially replace the electrons in equilibrating the charge.

\begin{figure}[h!]
  \centering
  \includegraphics[width=0.5\textwidth]{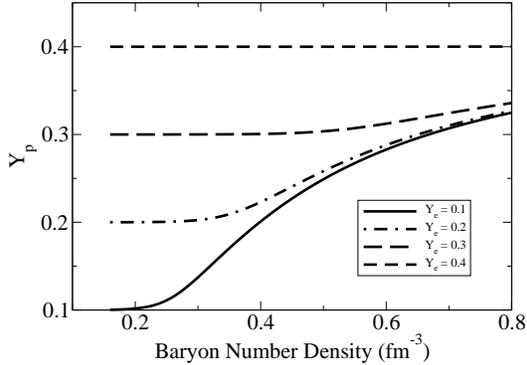}
  \caption{The proton fraction above nuclear saturation showing the effects of pions in the hot dense supernova environment. For small electron fractions more pions are created due to the dependence of the chemical potential on the isospin asymmetry parameter~$I$.}
 \label{fig:pion_prot_frac}
\end{figure}

\begin{figure}[h!]
  \centering
  \includegraphics[width=0.5\textwidth]{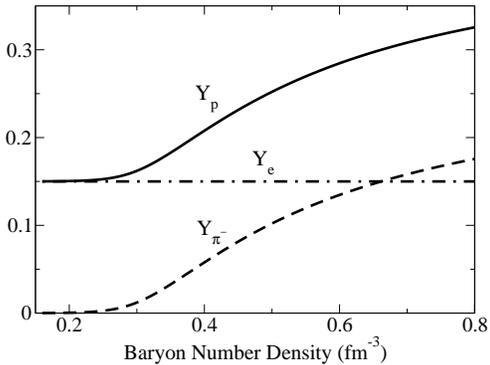}
  \caption{Pion charge fraction versus density at a temperature $T = 10$~MeV. At a density of about $n \approx 0.65 \text{ fm}^{-3}$, the pion charge fraction exceeds the electron fraction of the medium.}
 \label{fig:pion_charge_frac}
\end{figure}

From the solution to the pion chemical potentials [Eq.~(\ref{eqn:mupi})], one finds that as the density increases, negative pions are created due to the chemical potential constraints and the dispersion relation [Eq.~(\ref{eqn:dispersion})]. At the same time the number density of positively charged pions remains negligible due to the fact that it has a negative chemical potential.  Due to its dependence on both the symmetry energy [Eq.~(\ref{eqn:sym_energy})] and the isospin asymmetry parameter $I = (1-2Y_p)$, the pion chemical potential [Eq.~(\ref{eqn:mupi})] increases linearly with respect to density but decreases  linearly with respect to $Y_p$.  Therefore, for high electron fractions the pion chemical potential will remain small and charge equilibrium can be maintained solely among the electrons and protons.  

It should be noted, however, that as treated here, pions would not exist in the ground state configuration of a cold neutron star. As the temperature approaches zero the pionic effects diminish, until only the nucleon EoS contributes to the neutron star structure.


In the hot dense medium of supernovae, however, these pions tend to soften the hadronic EoS since they relieve some of the degeneracy pressure due to the electrons.  We have found that the reduction in pressure is relatively insensitive to the temperature of the medium and is lowered by approximately 10\% for all representative temperatures found in the supernova environment, as shown in Fig.~\ref{fig:pion_pressure}.  

\begin{figure}[h!]
  \centering
  \includegraphics[width=0.5\textwidth]{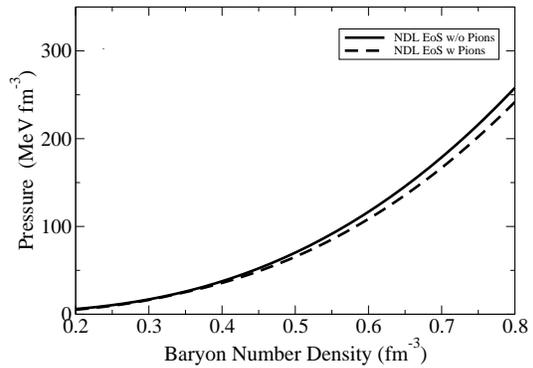}
  \caption{Pressure versus baryon number density showing a 10\% reduction in the pressure at high densities due to the presence of pions.}
 \label{fig:pion_pressure}
\end{figure}
This will affect SN core collapse models since it allows collapse to higher densities and temperatures without violating the requirement that the maximum neutron star mass exceed \mbox{$1.97 \pm 0.04 ~M_\odot$} for cold neutron stars.

\subsection{Hadron QGP Mixed Phase}
The constraint of global charge neutrality exploits the isospin restoring force experienced by the confined hadronic matter phase.  This portion of the mixed phase becomes more isospin symmetric than the pure phase because charge is transferred from the quark phase in equilibrium with it. 

Fig.~\ref{fig:isospin} shows the charge fractions of the mixed phase and hadronic phases. From this we see that the internal mixed phase region of a hot, proto-neutron star contains a positively charged region of nuclear matter and negatively charged regions of quark matter until a density of \mbox{$n_0 \sim 1.0~\text{fm}^{-3}$}. The presence of the isospin restoring force causes the thermal pionic contribution to the state variables to be negligible. This is due to the dependence of the pion chemical potential on the isospin asymmetry parameter $(1-2Y_p)$.  As the hadronic phase becomes more isospin symmetric, the pion chemical potential remains small compared to its effective mass. 

\begin{figure}[h!]
  \centering
  \includegraphics[width=0.5\textwidth]{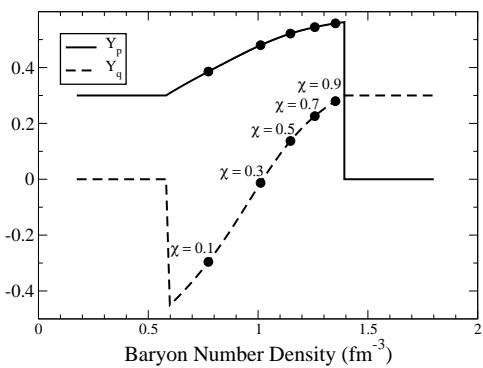}
  \caption{Charge fractions of the mixed quark and hadronic phase. Due to the redistribution of charge, the hadronic phase becomes isospin symmetric even exceeding $Y_p >$ 0.5. This has the effect of lowering the symmetry energy and thus reducing the pressure in the hadronic phase.}
 \label{fig:isospin}
\end{figure}

Since stars contain two conserved quantities, electric charge and baryon number, the coexistence region cannot be treated as a single substance, but must be evolved as a complex multicomponent fluid.  It is common in Nature to have global conservation laws and not necessarily locally conserved quantities. Hence, within the Gibbs construction the pressure is a monotonically increasing function of density. 

Fig.~\ref{fig:mixed_pressure} shows pressure versus density for various values of $Y_e$, through the mixed phase region into a phase of pure QGP.  One of the features shown is that as the density increases through the mixed region the slope of the pressure decreases slightly. This becomes more evident when the adiabatic index, $\Gamma$, is analyzed as a function of density as shown in Fig.~\ref{fig:gamma}. Here, we find that the EoS softens abruptly upon entering the mixed phase due to the fact that increasing density leads to more QGP rather than an increase in pressure. If this occurs while forming a proto-neutron star, its evolution will be affected as $\Gamma$ falls below the stability point of $\Gamma < 4/3$ for \mbox{$n \sim 1.1~\text{fm}^{-3}$}.  

\begin{figure}[h]
  \centering
  \includegraphics[width=0.5\textwidth]{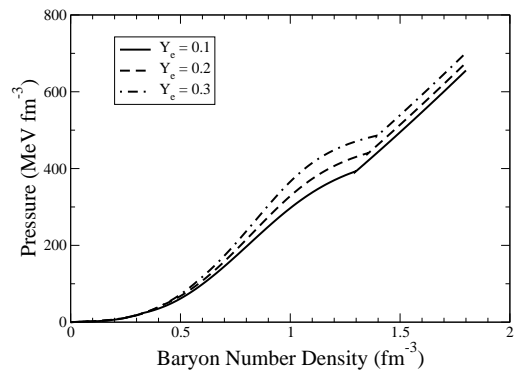}
  \caption{Pressure as a function of baryon number density through the mixed phase transition. The EoS softens significantly upon entering the mixed phase due to the larger number of available degrees of freedom.}
 \label{fig:mixed_pressure}
\end{figure}

The collapse simulations of~\cite{Fischer, Gentile} show that as $\Gamma$ falls below 4/3, a secondary core collapse ensues. The matter sharply stiffens upon entering the pure quark phase at \mbox{$n \sim 1.4 \text{ fm}^{-3}$}, and a secondary shock wave is generated. As this shock catches up to the initially stalled accretion shock, a more robust explosion ensues. 

\begin{figure}[h!]
  \centering
  \includegraphics[width=0.5\textwidth]{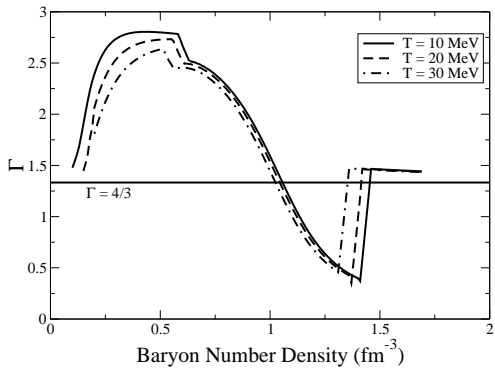}
  \caption{$\Gamma$ as a function of baryon number density showing the softening of the EoS as it enters the mixed phase regime. The EoS promptly stiffens as it exits the mixed phase into the pure quark matter phase due to losing the extra degrees of freedom supplied by the nucleons.}
 \label{fig:gamma}
\end{figure}

\begin{figure}[h!]
  \centering
  \includegraphics[width=0.5\textwidth]{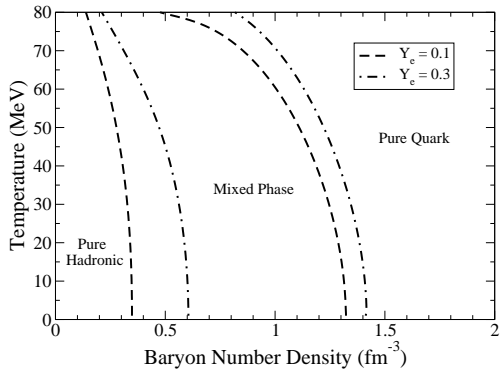}	
  \caption{Density-temperature phase diagram showing the density range of the mixed phase coexistence region for two values of $Y_e$ (dashed line for $Y_e=0.1$ and dash-dotted line for $Y_e=0.3$). The onset of the mixed phase is indicated by the set of curves on the left, while the curves on the right show the completion of the mixed phase. For low temperatures it is seen that the onset density is highly $Y_e$ dependent.}
 \label{fig:phase_plot}
\end{figure}


\begin{figure}[h]
  \centering
  \includegraphics[width=0.5\textwidth]{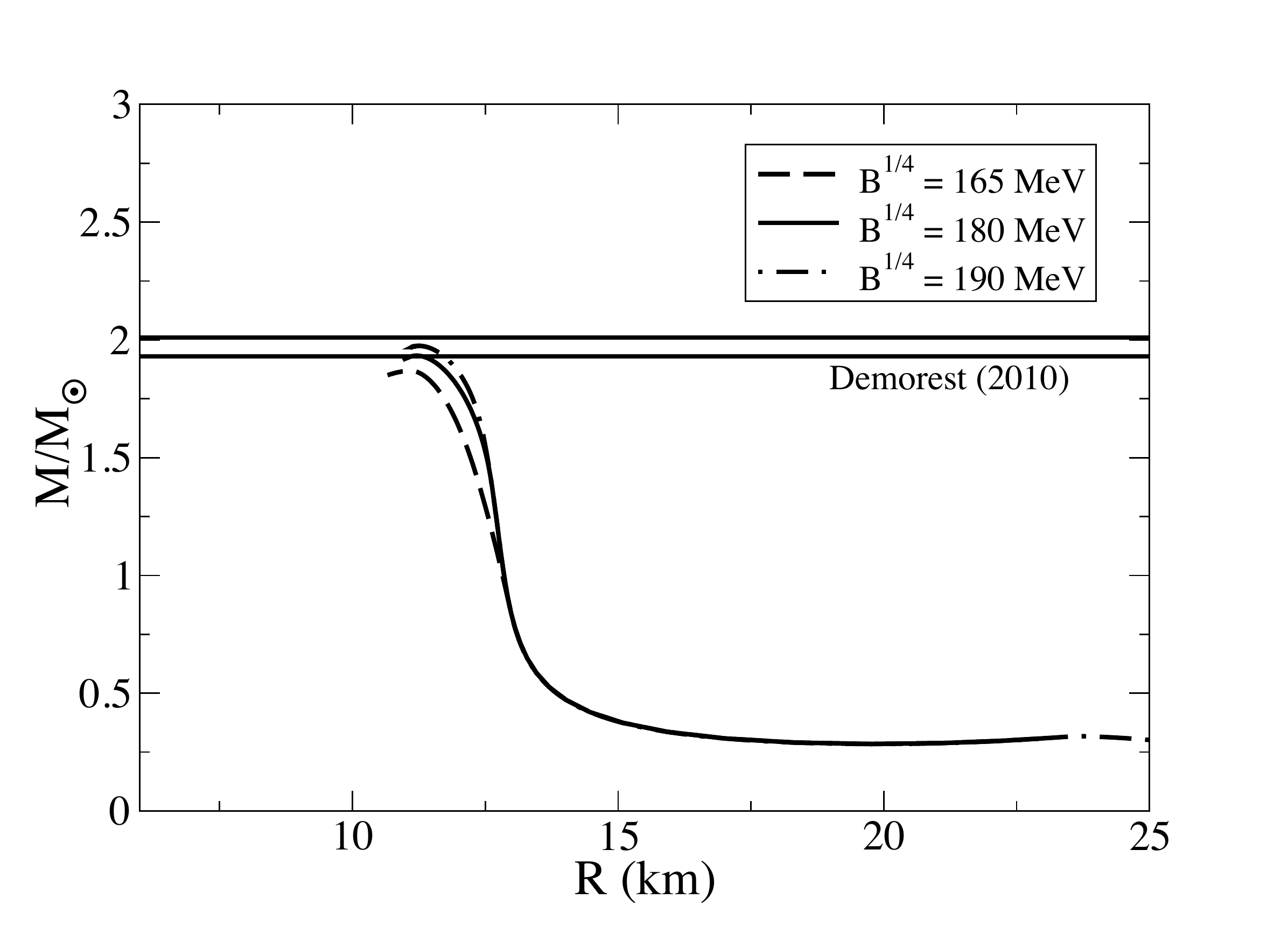}
  \caption{Neutron star mass-radius relation for various values of the bag constant B$^{1/4}$.  We find that a first order phase transition is consistent with the maximum mass neutron star measurement for our adopted value of B$^{1/4}$ = 180 MeV.}
 \label{fig:MaxMassBag}
\end{figure}

Another feature seen in Fig.~\ref{fig:mixed_pressure} is the $Y_e$ dependence of the onset density of the mixed phase. Fig.~\ref{fig:phase_plot} shows a phase diagram indicating the mixed phase transition temperature as a function of density for two values of $Y_e$. For higher temperatures the onset happens at lower densities as would be expected. However, for high electron fractions ($Y_e \sim 0.3$) such as those that can be found deep inside the cores of a proto-neutron star, the transition density remains quite high \mbox{$n_c \sim 0.6 ~\text{fm}^{-3}$}.  It is also of note that the coexistence region slightly decreases as the electron fraction is increased.

\subsection{Neutron Stars with QGP Interiors}
As stated previously the observation~\cite{Demorest} of a \mbox{$1.97 \pm 0.04 ~M_\odot$} neutron star has ruled out many exotic EoSs including many Hyperonic models~\cite{Lattimer2012}. However, using the range of bag constants determined by Eq.~(\ref{eqn:Tc}) we find that a first order phase transition to a QGP is consistent with the high maximum neutron star mass constraint~\cite{Demorest} for our fiducial NDL~EoS. In Fig.~\ref{fig:MaxMassBag}  we show that a bag constant \mbox{$B^{1/4} > 190$ MeV} is required to satisfy the maximum neutron star mass constraint. This imposes a low baryon density transition temperature of $T_c > 150$~MeV which is slightly below the current range of crossover temperatures determined from LGT~\cite{Kronfeld}.  Hence all allowed values of the Bag constant inferred from LGT are consistent with the neutron star mass constraint.  For our purpose we will adopt \mbox{$B^{1/4} = 190$ MeV} (corresponding to \mbox{$T_c \sim 150$~MeV}).
We note, however, that the maximum mass is not above the observational limit if you simply ignore the 2-loop corrections (while keeping the same choice for $B^{1/4}$).

Fig.~\ref{fig:MaxMass} compares the neutron star mass radius relation for the NDL~EoS for: 1) a hadronic EoS (solid line); 2) a first order QCD transition with $B^{1/4} = 190$~MeV (dot-dot dashed line); and 3) a simple QCD cross over transition (dotted line). Also, shown for comparison are results from the LS220 (dashed line), LS180 (dot-dash line) Shen EoS (long-dashed line) and the original Bowers \& Wilson EoS (dash-dash dotted line). Note, that all three versions of the NDL~EoS easily accommodate a maximum neutron star \mbox{mass $\ge 1.97 \pm 0.04 ~M_\odot$}.

\begin{widetext}
\begin{center}
\begin{figure}[hc]
  \centering
  \includegraphics[width=1.0\textwidth]{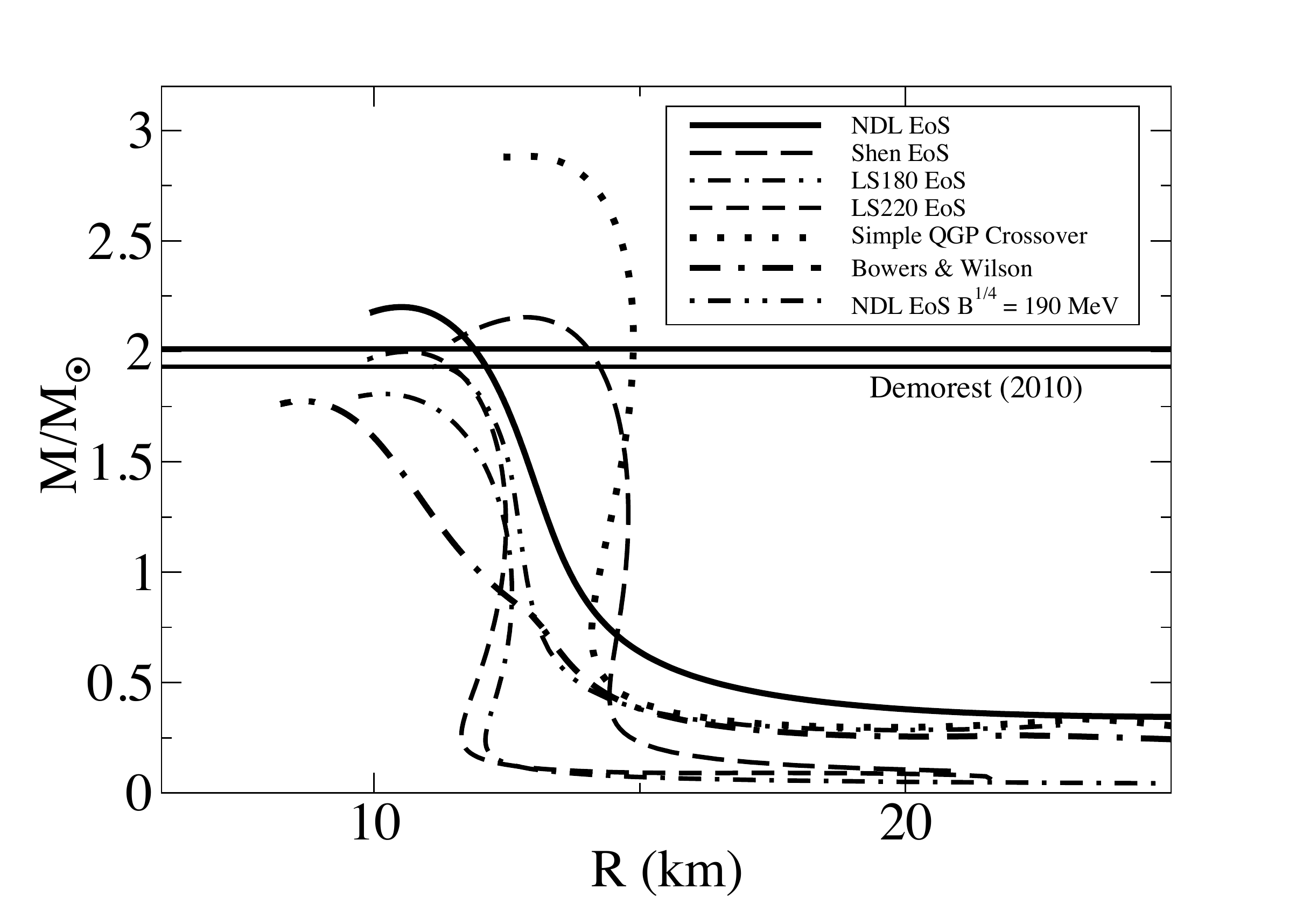}
  \caption{Mass-radius relation for the Shen (dash-dash-dashed line), Lattimer~\&~Swesty 180 \& 220 (dashed line and dot-dash-dot-dash), Bowers~\&~Wilson EoS (dash-dash dotted line), the NDL~EoS with (solid line) and without (solid line) a mixed phase transition to quark gluon plasma as well as a simple crossover transition (dotted line) to a QGP. Note that all of these curves satisfy the \mbox{$1.97 \pm 0.04$ M$_\odot$} astrophysical constraint except for the Bowers \& Wilson and LS180. }
 \label{fig:MaxMass}
\end{figure}
\end{center}
\end{widetext}

\section{Conclusion}
We have discussed a much updated and improved equation of state based upon the original Livermore framework~\cite{Bowers82, WilsonMathews}. We have shown that it is complementary to the most frequently employed equations of state for core-collapse supernovae due to Shen~et.~al.~\cite{Shen98a, Shen98b,Shen2011}, and Lattimer~\&~Swesty~\cite{LS91}.  This NDL~EoS is consistent with the known constraints of symmetric nuclear matter and observed properties of neutron stars and pulsars, whereas the previous version~\cite{Bowers82,WilsonMathews} was not. We found that consistently applying the constraints on symmetric nuclear matter, combined with the observation of a \mbox{$1.97 \pm 0.04 ~M_\odot$} neutron-star, places a stronger limitation on the Skewness coefficient $Q_0$ than is available in the literature. A first order phase transition to a QGP phase was also discussed in the context of a Gibbs construction. Applying the constraints from LGT for the range of low-baryon-density crossover temperatures, we were able to match the known constraints of the current maximum neutron star mass measurement for a bag constant \mbox{$B^{1/4} \ge 140$ MeV}. On the other hand, if there is a cross over QCD transition the neutron star mass constraint can be easily accommodated for any value of the bag constant.

 This confirms that a core collapse explosion paradigm, including a transition to quark gluon plasma may impact the neutrino light curve, shock dynamics, and heavy element nucleosynthesis via the $\nu$-process and $\nu p$ process both in supernovae and/or black hole formation. The consequences of this new NDL~EoS for the dynamics of core collapse supernovae, along with its impact on nucleosynthesis, will be explored in forthcoming manuscripts.

\begin{acknowledgements}
Work at the University of Notre Dame is supported by the U.S. Department of Energy under Nuclear Theory Grant DE-FG02-95-ER40934. One of the authors (N.Q.L.) was supported in part by the National Science Foundation through the Joint Institute for Nuclear Theory (JINA).
\end{acknowledgements}

\bibliography{References.bib}

\end{document}